\documentclass[10pt,a4paper]{article}
\usepackage{a4wide}
\usepackage{amsfonts}
\usepackage{amsmath}
\usepackage{color}
\usepackage{environ}
\usepackage{graphicx}
\usepackage{times}
\usepackage{url}

\newcommand{\eqlabel}[1]{\label{eq:#1}}
\newcommand{\eq}[1]{\eqref{eq:#1}}
\def\eqreftwo(#1,#2){(\ref{eq:#1},\ref{eq:#2})}
\newcommand{\eqtwo}[1]{\eqreftwo(#1)}
\NewEnviron{eqsplit}[1]{\begin{align}\eqlabel{#1}\begin{split} \BODY \end{split}\end{align}}

\newcommand{\secn}[1]{Section~\ref{sec:#1}}
\newcommand{\seclabel}[1]{\label{sec:#1}}

\newcommand{\fig}[1]{fig.~\ref{fig:#1}}
\newcommand{\figs}[1]{figures~\ref{fig:#1}}
\newcommand{\figref}[1]{\ref{fig:#1}}
\newcommand{\sglfigure}[3]{
  \begin{figure}[htbp]
  \centerline{\includegraphics{#1}}
  \caption[]{#2}
  \label{fig:#3}
  \end{figure}
}

\renewcommand{\@}{\partial}        
\newcommand{\const}{\mathrm{const}} 
\newcommand{\D}[1]{\cdot10^{#1}}
\renewcommand{\d}{\mathrm{d}}      
\renewcommand{\deg}{\circ}      
\newcommand{\df}[2]{\frac{\partial #1}{\partial #2}}
\newcommand{\diag}{\mathrm{diag}}       
\DeclareMathOperator{\Heav}{H}     
\renewcommand{\i}{i}               
\newcommand{\Mx}[1]{\left[\begin{array}{ccccccc}#1\end{array}\right]}
\newcommand{\mx}[1]{\mathbf{#1}}
\providecommand{\norm}[1]{\left\lVert#1\right\rVert}
\renewcommand{\O}[1]{\mathcal{O}\left(#1\right)}
\newcommand{\Real}{\mathbb{R}}
\newcommand{\T}{^{\!\top}}
\newcommand{\takenat}[2]{\left.#1\rule[-2ex]{0pt}{4ex}\right|_{#2}}

\newcommand{\+}[2]{\def#1{#2}}

\+{\Amp}{\u_s}                
\+{\bDuu}{\mx{D_{uu}}}        
\+{\bDuup}{\mx{D'_{uu}}}      
\+{\bDuus}{\mx{D''_{uu}}}     
\+{\bDuut}{\mx{D'''_{uu}}}    
\+{\bDuuq}{\mx{D^{\mathrm{iv}}_{uu}}}       
\+{\bQuu}{\mx{Q_{uu}}}        
\+{\bDuv}{\mx{D_{uv}}}        
\+{\bDvu}{\mx{D_{vu}}}        
\+{\bDvv}{\mx{D_{vv}}}        
\+{\bfu}{\mx{f_u}}            
\+{\bfuf}{\mx{f'_u}}          
\+{\bFu}{\mx{F_u}}            
\+{\bfv}{\mx{f_v}}            
\+{\bFv}{\mx{F_v}}            
\+{\bg}{\mx{g}}               
\+{\bG}{\mx{G}}               
\+{\bu}{\mx{u}}               
\+{\bv}{\mx{v}}               
\+{\bvp}{\tilde{\bv}}         
\+{\Du}{D_u}                  
\+{\Duu}{D_{uu}}              
\+{\Duv}{D_{uv}}              
\+{\Dv}{D_v}                  
\+{\Dvu}{D_{vu}}              
\+{\Dvv}{D_{vv}}              
\+{\Dw}{D_w}                  
\+{\Dwp}{D'_w}                
\+{\dt}{\Delta_t}             
\+{\dx}{\Delta_x}             
\+{\deps}{\mu}                
\+{\eps}{\epsilon}            
\+{\F}{F}                     
\+{\f}{f}                     
\+{\fna}{a}                   
\+{\fne}{\varepsilon}         
\+{\fnk}{k_1}                 
\+{\fnal}{\alpha}             
\+{\fnbe}{\beta}              
\+{\fnga}{\gamma}             
\+{\fnde}{\delta}             
\+{\G}{G}                     
\+{\g}{g}                     
\+{\icw}{\sigma}              
\+{\k}{k}                     
\+{\L}{L}                     
\+{\m}{m}                     
\+{\n}{n}                     
\+{\t}{t}                     
\+{\u}{u}                     
\+{\v}{v}                     
\+{\w}{w}                     
\+{\x}{x}                     
\+{\xc}{x_c}                  
\+{\z}{z}                     

\begin{document}

\title{
Quasisolitons in self-diffusive excitable systems, or \\
Why asymmetric diffusivity obeys the Second Law
}

\author{%
  V. N. Biktashev$^1$, %
  M. A. Tsyganov$^2$ \\
  $^1$  College of Engineering, Mathematics and Physical Sciences, \\
  University of Exeter, Exeter EX4 4QF, UK, and \\
  EPSRC Centre for Predictive Modelling in Healthcare, \\
  University of Exeter, Exeter, EX4 4QJ, UK \\
  $^2$ Institute of Theoretical and Experimental Biophysics, \\
  Pushchino, Moscow Region, 142290, Russia 
}%
\date{\today}
\maketitle

\begin{abstract}
  Solitons, defined as nonlinear waves which can reflect from
  boundaries or transmit through each other, are found in
  conservative, fully integrable systems. Similar phenomena, dubbed
  quasi-solitons, have been observed also in dissipative,
  ``excitable'' systems, either at finely tuned parameters (near a
  bifurcation) or in systems with cross-diffusion. Here we demonstrate
  that quasi-solitons can be robustly observed in excitable systems
  with excitable kinetics and with self-diffusion only. This includes
  quasi-solitons of fixed shape (like KdV solitons) or envelope
  quasi-solitons (like NLS solitons).  This can happen in systems with
  more than two components, and can be explained by effective
  cross-diffusion, which emerges via adiabatic elimination of a fast
  but diffusing component. We describe here a reduction procedure can
  be used for the search of complicated wave regimes in
  multi-component, stiff systems by studying simplified, soft
  systems.%
\end{abstract}


\section{Motivation}
\seclabel{introduction}
Reaction-diffusion systems are an important mathematical
tool in studying dissipative structures. For the properties of
solutions of these systems, diffusion is as important as is reaction.
For instance, Turing structures are possible only if there is
significant difference in the diffusion coefficients between activator
and inhibitor species~\cite{Turing-1952}.  Most of literature on waves
and patterns is focussed on self-diffusion, when the flux of a
reacting species is defined by the gradient of that same
species. However, this ignores the phenomenon of cross-diffusion, when
the flux of one species depends on gradient of another species.

There is a large variety of interesting regimes in systems with
nonlinear kinetics and cross-diffusion instead of or in additional to
self-diffusion~\cite{%
  Tsyganov-Biktashev-2014%
}. Some of these regimes present a considerable theoretical interest,
since they manifest properties that are traditionally associated with
very different ``realms'': on one hand, these are waves that preserve
stable and often unique profile, speed and amplitude, similar to nerve
pulse, on the other hand, they can penetrate through each other or
reflect from boundaries, such as waves in linear systems or solitons
in conservative nonlinear systems. It is interesting where in nature
such regimes can be observed.

Originally, these regimes have been discovered in models motivated by
population dynamics, where (nonlinear) cross-diffusion appears
naturally as taxis of micro- or macro-organisms with respect to each
other's population density~\cite{%
  Tsyganov-etal-2003,%
  Biktashev-Tsyganov-2009%
}. There are also examples of similar models
occurring in physical applications, see
e.g.~\cite{Cartwright-etal-1997}. However, real cross-diffusion is
actually observed in chemical solutions along with the
self-diffusion~\cite{Vanag-Epstein-2009}, which brings up the
question of the possibility of observing the quasi-solitons in
excitable chemical reactions, such as BZ reaction or some of its
modifications, which would be a convenient experimental object to
observe quasi-solitons. 

This question encounters a number of obstacles due to physical
constraints. Firstly, the cross-diffusion must always be
nonlinear in order to preserve positivity of chemical
concentrations~\cite{Gorban-etal-2011}. Then, there is a fundamental
requirement that the diffusivity matrix should have all real and
positive eigenvalues~\cite{Kirkaldy-etal-1963}, which follows from
Onsager reciprocal relations~\cite{Onsager-Fuoss-1932,Nobel-1968}
based on the Second Law of Thermodynamics.
The first of these obstacles does not seem particularly essential:
quasi-solitons have been observed in models both with
linear~\cite{%
  Biktashev-Tsyganov-2005,%
  Biktashev-Tsyganov-2011,%
  Tsyganov-Biktashev-2014%
} and non-linear~\cite{%
  Tsyganov-etal-2003,%
  Biktashev-Tsyganov-2009%
} cross-diffusion. The second obstacle is more serious, as in all the
known theoretical examples, the diffusivity matrix required for
observation of quasi-solitons, has zero or complex eigenvalues.
Hence, as argued e.g. in~\cite[page 899]{Vanag-Epstein-2009}, it would
appear that there are no chances to observe these regimes in chemical
excitable systems.

In the present paper, we seek to look into this question a bit deeper.
We are led by the observation that the above fundamental requirement
is about actual diffusivities, which would appear in reactions written
in all their elementary steps, for all intermediate reacting species,
whereas for studies of waves and pattern formation one typically uses
reduced mathematical models, and the
diffusivities in them are \emph{effective}. Pattern-forming reactions
are supposed to be supported by a source of energy, say in the form of
constant supply of a reagent which is consumed in the reaction, whereas
the Second Law applies to closed systems. Although this source of
energy (and of ``negative entropy'', in Schr\"odinger's
sense~\cite{Schroedinger-1944}) is in the reaction part, 
in reduced models this part is actually related to
the diffusion part.
In reduced models, several intermediate stages
and species are lumped together or ``adiabatically eliminated'', say
by asymptotic methods. In classical approaches, this
is typically done for well-mixed reaction 
(see e.g. the review~\cite{Radulescu-etal-2012}), whereas those
intermediate species may, and often are,
diffusive. Reduction in the full reaction-diffusion context, with
account of different diffusivities of the eliminated intermediate
reagents is not often done; however see~\cite{Kuznetsov-etal-1994} for
an example. That example is for an
ecological model, but the mathematics involved 
is equally applicable for chemical reactions. That example shows,
firstly, that effective diffusion coefficients occurring after the
elimination are significantly different from those before (in
particular, cross-diffusion terms appear where have been none
originally) and, secondly, that these effective diffusion coefficients
depend on the reaction kinetics linking the eliminated species with
the main species --- which is the place through which the remoteness
from the thermodynamic equilibrium creeps right into the diffusion
matrix.  Another example, more specifically about chemical systems, is
the complex Ginzburg-Landau equation, which is in fact a
reaction-diffusion system where the diffusion matrix has a complex
pair of eigenvalues, but it can be obtained by asymptotic reduction
from a reaction-diffusion system with a diagonal diffusion matrix; see
e.g.~\cite[Appendix B]{Kuramoto-1980}.

In this paper, we illustrate how the fast-slow analysis (adiabatic
elimination of fast variables) in the spatially-extended context can
be used to reconstruct full (fuller) reaction-diffusion systems with
purely self-diffusing reagents, from reduced reaction-diffusion
systems with cross-diffusion. We apply this procedure to produce
examples of self-diffusion systems that manifest quasi-soliton
solutions. Since the fast-slow analysis is an asymptotic procedure,
i.e. is formally valid in certain parametric limits but has to be
applied for fixed parameter values, it inherently has limited
accuracy. The procedure we exploit here is valid when the spatial
scale of the solutions is large compared to the diffusion length of
the fast species, and the reduced equations become non-local when this
is not fulfilled. To relax this restriction, we suggest a euristic
procedure which enhances the accuracy of the reduction/reconstruction for particular
wave-like solutions.

The structure of the paper is as follows. In~\secn{general}
we discuss the process of adiabatic elimination of a fast reacting
species in the spatially-extended context, and show how the
diffusivity matrix may be affected by that
elimination. In~\secn{simple-example} we present a simple application
of this theory: namely, by ``working backwards'' the adiabatic
elimination, we construct a three-component reaction-diffusion system
with self-diffusion only, which corresponds to a two-component system
with cross-diffusion and quasi-soliton solutions of various
kinds. 
We conclude by discussing
possible implications of our findings for future research.

\section{Methods: Fast-slow reduction for reaction-diffusion systems}
\seclabel{general}

Our asymptotic argument is based on two assumptions. Firstly, we 
consider a reaction-diffusion system, which has a fast and a slow
subsystem,
\begin{align}
  \@_{\t}{\bu} &= \bfu(\bu,\bv) + \bDuu\nabla^2\bu + \bDuv\nabla^2\bv,
                                        \eqlabel{ueq} 
  \\
  \eps\@_{\t}{\bv} &= \bfv(\bu,\bv) + \bDvu\nabla^2\bu +
                     \bDvv\nabla^2\bv , 
                                        \eqlabel{veq}
\end{align}
where $\bu,\bfu\in\Real^\m$, $\bv,\bfv\in\Real^\n$,
$\bDuu\in\Real^{\m\times\m}$,
$\bDuv\in\Real^{\m\times\n}$,
$\bDvu\in\Real^{\n\times\m}$,
$\bDvv\in\Real^{\n\times\n}$,
and the parameter 
$0<\eps\ll1$ represents the time separation between the slow subsystem
$\bu$ and the fast subsystem $\bv$. 
We assume that the fast subsystem~\eq{veq} has a unique globally
stable equilibrium for $\bv=\bg(\bu)$ at any fixed $\bu$ of interest,
\begin{align}
  \bfv(\bu,\bg(\bu))\equiv 0, 
  \quad
  \bFv(\bu)\equiv\takenat{\left(\df{\bfv(\bu,\bv)}{\bv}\right)}{\bv=\bg(\bu)}<0. \eqlabel{vinv}
\end{align}
The second assumption, to be clarified and formalized later, is that
the solutions of interest are sufficiently smooth in space. 

Linearizing equation \eq{veq} around the stable equilibrium, 
\begin{align*}
  \bv=\bg(\bu)+\bvp,
  \quad
  \norm{\bvp}\ll 1,
\end{align*}
and
considering the limit $\eps\searrow0$, we have
\begin{align}
  \left(\bFv(\bu) + \bDvv\nabla^2\right) \bvp 
  \approx 
  - \bDvu \nabla^2 \bu
  - \bDvv\nabla^2\bg(\bu)
  . \eqlabel{bvpeq}
\end{align}
At this point we see that our solutions should be so smooth that
$\nabla^2 \bu$ is small and therefore the resulting $\bvp$ is
small, justifying the linearization made. Formally, we set
$\nabla^2 \bu=\O{\deps}$, $\bvp=\O{\deps}$ where $\deps$ is a
(symbolical) small parameter. 
Then we can formally resolve equation \eq{bvpeq}, keeping terms up to
and including
$\O{\deps^2}$, to get
\begin{align*} &
  \bvp \approx 
  \left( - \bFv^{-1} + \bFv^{-1}\bDvv\bFv^{-1}\nabla^2 \right) 
  \times\nonumber\\&\times
  \left(
    \bDvu \nabla^2 \bu
    + \bDvv\nabla^2\bg(\bu)
  \right) +\O{\deps^3} .
\end{align*}
Substitution of the thus found $\bv$ into equation \eq{ueq} leads to
\begin{align}
    \@_{\t}{\bu} = \bfuf(\bu) 
    + \bDuup(\bu) \nabla^2\bu 
    + \bDuus(\bu) \nabla^2\bg(\bu)
    + \O{\deps^2},                  \eqlabel{first-order}
\end{align}
where %
$\bfuf(\bu)=\bfu(\bu,\bg(\bu))$, %
$\bFu=\takenat{\@_{\bv}\bfu(\bu,\bv)}{\bv=\bg(\bu)}$, and %
\begin{align*}
& \bDuup=\bDuu - \bFu \bFv^{-1} \bDvu, \\
& \bDuus(\bu)=\bDuv - \bFu\bFv^{-1}\bDvv.
\end{align*}
We see that the diffusion in the reduced system~\eq{first-order} is not only different
from the full system, but is typically nonlinear even if the
original diffusion was fully linear.

In the system~\eq{first-order} we have dropped $\O{\deps^2}$ terms as they are
complicated and will not be needed in this paper in the general form.
In the special case when the fast subsystem is linear and is linked to
the slow system in a linear way, the situation simplifies. We then
have $\bg(\bu)=\bG\bu$, $\bG=\const$, and
\begin{align}
  \@_{\t}{\bu} = \bfuf(\bu) 
  + \bDuut \nabla^2\bu 
  + \bQuu \nabla^4\bu 
  + \O{\deps^3},                    \eqlabel{second-order}
\end{align}
were
\begin{align}
  \bDuut =
  \bDuu - \bFu\bFv^{-1}\bDvu
    + \bDuv\bG - \bFu\bFv^{-1}\bDvv\bG
\end{align}
and
\begin{align}
  \bQuu= 
    \left(
      \bFu\bFv^{-1}\bDvv
      - \bDuv
    \right)\bFv^{-1}\left(
      \bDvu + 
      \bDvv\bG
      \right) .
\end{align}
The second-order approximation~\eq{second-order} contains a biharmonic
operator. 
We can
approximately replace it with a reaction-diffusion system for
solutions which are approximately oscillatory in space
with one dominant wavenumber $\k$, so that
\begin{align*}
  \nabla^4\bu \approx -\k^2\nabla^2\bu .
\end{align*}
Then we have 
\begin{align}
  \@_{\t}{\bu} = \bfuf(\bu) 
  + \bDuuq \nabla^2\bu                  \eqlabel{second-order-qh}
\end{align}
where
$\bDuuq \approx \bDuut - \k^2 \bQuu$.
If the solutions of interest, or their Laplacians, are not approximate
solutions of the Helmholtz equation, this second-order
approximation is clearly only a euristic.

The direct transformation from system~\eqtwo{ueq,veq} to
system~\eq{first-order} or~\eq{second-order-qh}, and its inverse, can
be used for searching for nontrivial regimes in the full system, based
on the existing experience of nontrivial regimes in the analogues of
the reduced systems:
\begin{enumerate}
\item Do parametric search in the reduced system, which has fewer
  parameters and often is easier to compute, as it is less stiff;
\item Once an interesting regime is found, estimate the dominant
  wavenumber $\k$ for it, say as the maximum of the power spectrum of
  the spatial profile;
\item Use the inverse transformation to obtain parameters for the full
  system corresponding to the found parameters of the reduced system;
\item See what solutions the full system with these parameters will
  produce.
\end{enumerate}

\section{Results}
\seclabel{simple-example}
\subsection{Direct transform (reduction)}

As a simple example, we now consider a three-component extension of a
two-component system with nonlinear kinetics, where a third component
is added, which has fast linear kinetics and linked to the other
components in a linear way:
\begin{eqsplit}{three-comp}
    \df{\u}{\t} &= \f(\u,\v) + \fnal \w + \Du \nabla^2\u, \\
    \df{\v}{\t} &= \g(\u,\v) + \fnbe \w + \Dv \nabla^2\v, \\
    \eps\df{\w}{\t} &= \fnga\u + \fnde \v - \w + \Dw \nabla^2\w . 
\end{eqsplit}
In terms of equations~\eqtwo{ueq,veq}, this means
\begin{align*} &
    \bu=\Mx{\u,\v}\T,
    \quad
    \bv=\Mx{\w},
    \\&
    \bfu=\Mx{\f+\fnal\w,\g+\fnbe\w}\T,
    \\&
    \bfv=\Mx{\fnga\u + \fnde \v - \w},
    \\&
    \bDuu=\diag\left(\Du,\Dv\right),
    \quad
    \bDvv=\Mx{\Dw},
    \\&
    \bDuv=\Mx{0,0}\T,
    \quad
    \bDvu=\Mx{0,0}
\end{align*}
which leads to the reduced system~\eq{second-order-qh} in the form
\begin{eqsplit}{two-comp}
  \@_{\t}{\u} &= \F(\u,\v) + \Duu\nabla^2\u + \Duv\nabla^2\v , \\
  \@_{\t}{\v} &= \G(\u,\v) + \Dvu\nabla^2\u + \Dvv\nabla^2\v ,
\end{eqsplit}
where
\begin{eqsplit}{direct-transform}
  & \F(\u,v) = \f(\u,\v) + \fnal(\fnga\u + \fnde \v), \\
  & \G(\u,\v) = \g(\u,\v) + \fnbe(\fnga\u + \fnde \v), \\
  & \Duu = \Du + \fnal\fnga\Dwp, \\
  & \Duv = \fnal\fnde\Dwp, \\
  & \Dvu = \fnbe\fnga\Dwp , \\
  & \Dvv = \Dv+\fnbe\fnde\Dwp ,
\end{eqsplit}
and
\begin{align}
  \Dwp = \Dw(1 - \k^2 \Dw) .            \eqlabel{Dwp}
\end{align}
It is easy to see that the resulting diffusivity matrix
\begin{align*}
  \bDuuq = 
  \Mx{\Du&0\\0&\Dv}+\Dwp\Mx{\fnal\\\fnbe}\Mx{\fnga&\fnde},
\end{align*}
depends on the reaction kinetics parameters and does not have to have
real eigenvalues or even be positive definite (although the utility of
the reduced model is of course stipulated by the well-posedness of its
Cauchy problems).

\subsection{Inverse transform (reconstruction)}

Equations~\eq{direct-transform} allow to determine parameters of the
two-component system~\eq{two-comp} from the given
system~\eq{three-comp}. For the study presented later, we need also to
be able to do the opposite: assuming that a 
two-component system~\eq{two-comp} is known, and is known to be a reduction of a
three-component system~\eq{three-comp}, to reconstruct the parameters
of that three-component system. One obvious restriction is that it is
not possible to reconstruct the value of parameter $\eps$ since the
known system~\eq{two-comp} corresponds to the zero limit of that
parameter. However, the non-uniqueness of the inverse transform is
even stronger than that, and there are infinitely many ways to choose
parameters of~\eq{three-comp}, which would correspond to the
same~\eq{two-comp}, even disregarding the variety of $\eps$. For instance,
we can also arbitrarily fix values of $\fnga$, $\Dw$, $\Dv$, and for
given parameters of~\eq{two-comp} (and given dominant wavenumber $\k$,
if using the euristic second-order approximation), 
obtain the remaining parameters of~\eq{three-comp} via
\begin{align*}
  & \fnal = \frac{\Duv}{\fnde\Dwp}, \\
  & \fnbe = \frac{\Dvu}{\fnga\Dwp}, \\
  & \fnde = \fnga\frac{\Dvv-\Dv}{\Dvu}, \\
  & \Du = \Duu - \frac{\Duv\Dvu}{\Dvv-\Dv}, \\
  & \f(\u,v) = \F(\u,\v) - \fnal(\fnga\u + \fnde \v), \\
  & \g(\u,\v) = \G(\u,\v) - \fnbe(\fnga\u + \fnde \v), 
\end{align*}
and $\Dwp$ is still related to $\Dw$ via equation~\eq{Dwp}.
The only restriction is that the resulting reconstructed system should
be physically reasonable. This, in particular, requires that
$\Du\ge0$.
The latter will be satified if $\Dv$ is chosen so that
either $\Dv\le\Duu\Dvv/(\Du+\Duu)$ or $\Dv\ge\Dvv$.

\subsection{Quasi-solitons in reduced system with FitzHugh-Nagumo kinetic}
\seclabel{results}

Numerical illustrations of some nontrivial regimes obtained in the
two-component system, and corresponding nontrivial regimes found in
the reconstructed three-component system are presented
in~\figs{fh2b-prof} and~\figref{fhdp}. These are for the two-component
system~\eq{two-comp} with the FitzHigh-Nagumo kinetics, taken in the
form 
\begin{eqsplit}{FHN}
  \F & = \u(\u-\fna)(1-\u) - \fnk\v, \\
  \G & = \fne \u,
\end{eqsplit}
and the corresponding three-component system~\eq{three-comp}. We used the
parameter values $\fne=0.01$, $\fnk=10$, $\eps=5.5\D{-4}$,
$\Duu=\Dvv=0.025$, $\Duv=1$, $\Dvu=-1$, $\fnga=30$, $\Dv=0.001$,
$\Dw=0.04$, and a selection of values of $\fna$ as shown to the left
of the panels in~\fig{fh2b-prof}.  The dominant wavenumbers $\k$ for
the inverse transform were obtained as the position of the maximum of
the power spectrum of the signal $\z(\x)=\u(\x,\t)+\i\v(\x,\t)$ of a
selected solution $(\u,\v)$ of the two-component system, taken at a
selected fixed $\t$. Specifically, we had $\k=0.864$ for $\fna=0.07$,
$\k=0.879$ for $\fna=0.25$ and $\k=0.942$ for $\fna=0.35$.
For simulations, we used the same numerical methods as those described
in~\cite{Biktashev-Tsyganov-2011,Tsyganov-Biktashev-2014}.  Computations were done using
second-order space differencing on the uniform grid with space step
$\dx=0.1$. Time differencing was first-order, explicit in the reaction
terms and implicit in the cross-diffusion terms, with time step
$\dt=2.5\D{-4}$ for the two-component system, and fully explicit with
time step $\dt=2\D{-5}$ for the three-component system.  The space
interval was $\x\in[0,\L]$. %
  Initial conditions were set as $\u(\x,0) = \Amp\Heav(\icw-\x)$,
  $\v(\x,0)=0$, $\w(\x,0)=0$ to initiate a wave starting from the left
  end of the domain. Here $\Heav()$ is the Heaviside function, the
  wave seed amplitude was typically chosen as $\Amp=2$ and length as
  $\icw=4$.
  To simulate propagation ``on an infinite line'', $\L=\infty$, for
  \fig{fh2b-prof}, we instantanously translated the solution by
  $\delta\x_2=20$ away from the boundary each time the pulse, as
  measured at the level $\u=0.1$, approached the boundary to a
  distance smaller than $\delta\x_1=40$, and filled in the new
  interval of $x$ values by extending the $u$, $v$ and $w$ variables
  at constant levels. %
The profiles in~\fig{fh2b-prof} are drawn in moving
  frames of reference; the ``co-moving'' space coordinate is defined as
  $\x-\xc$, where $\xc=\xc(\t)$ is the center of mass of $\v^2$ at the
  time moment $\t$, that is,
  $\xc(\t)=\int_0^\L\x\v(\x,\t)^2\,\d\x\big/\int_0^\L\v(\x,\t)^2\,\d\x$. 

\sglfigure{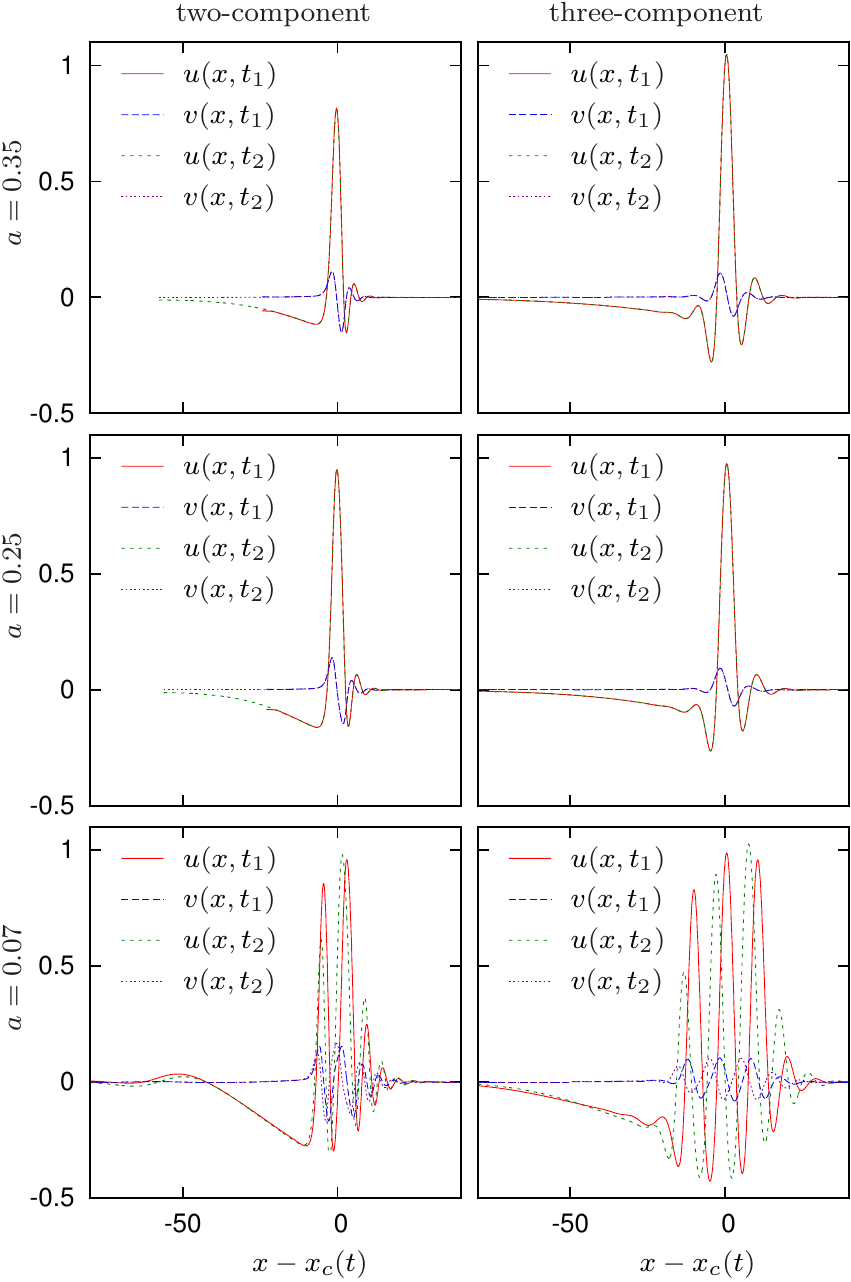}{%
  Selected profiles of quasi-soliton solutions for the FitzHugh-Nagumo
  kinetics, in co-moving frames of reference.
  Left column: two-component model with self- and
  cross-diffusion. Right column: corresponding reconstructed
  three-component model with self-diffusion only. The values of the
  parameter $\fna$ are given to the left of the panels, for other
  parameters see the text. The coordinate $\x-\xc$ is 
    in the comoving
  frame of reference, see text for detail. For each case, two
  profiles are shown at the moments $\t=\t_1$ and $\t=\t_2$ separated
  by an interval $\t_2-\t_1=150$; however, the consecutive profiles
  are only different for $\fna=0.07$ case, as in the other two cases,
  the waves have steady shapes %
  (see also Apple QuickTime movies corresponding to these regimes in
  Supplementary information). %
}{fh2b-prof}

\sglfigure{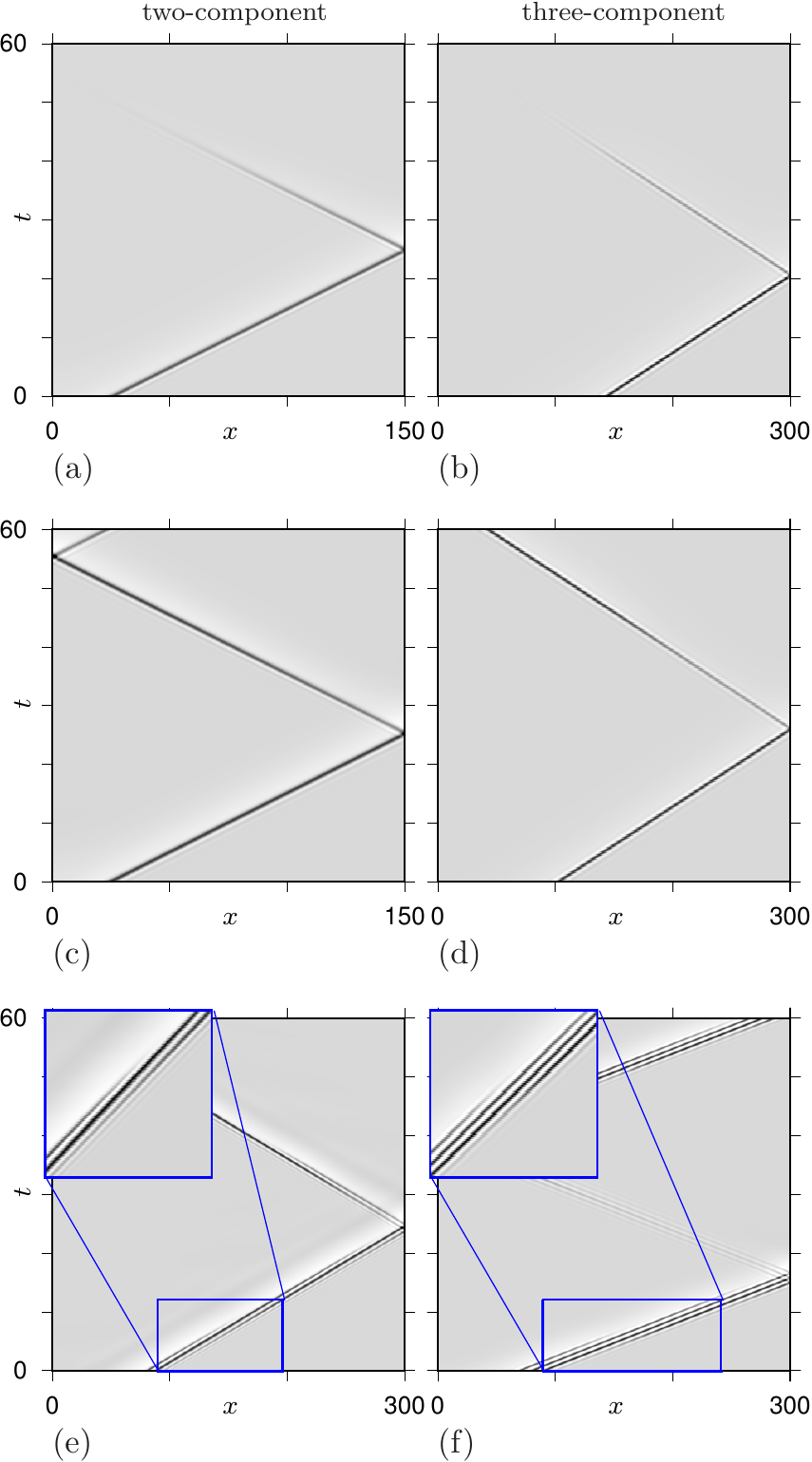}{%
  Density plots of the solutions represented in~\fig{fh2b-prof}, in
  particular: (a), (b) correspond to $\fna=0.35$, (c), (d) correspond
  to $\fna=0.25$, (e), (f) correspond to $\fna=0.07$, left column
  represents two-component model with cross-diffusion, right column
  the corresponding reconstructed three-component model with
  self-diffusion only. White corresponds to $\u=-0.3$, black
  corresponds to $\u=1.1$. The insets in panels (e), (f)
    (designated by solid blue lines) show selected fragments of the
    density plots magnified, to reveal the fine structure of group
    quasi-solitons %
  (see also Apple QuickTime movies corresponding to these regimes in
  Supplementary information). %
}{fhdp}

For $\fna=0.35$ and $\fna=0.25$, both 2-component and 3-component
systems support propagation of a solitary pulse with oscillating front
and monotonous back. These shapes are steady: as one can
  see in the top four panels of~\fig{fh2b-prof}, in the moving frame of
  reference, the solutions do not depend on time.
The difference is that for
$\fna=0.35$, the pulses annihilate upon collision with a bondary, ---
or, rather, reflect but lose their magnitude, cannot
recover and eventually decay. Whereas for $\fna=0.25$, the
pulses in both systems successfully reflect from boundaries. One can
see in~\fig{fhdp}(c) the reflection from the right boundary, followed
by reflection from the left boundary; the process repeats after that
(not shown). The behaviour in the three-component variant of the
system is similar, although the second reflection is not shown in
panel (d). %
  We emphasize here that the established shape, speed and amplitude of
  propagating pulses depend only on the parameters of the model, but
  not on initial conditions, as long as the initial condition is such
  that a propagating pulse is initiated. For instance, the profiles of
  the solutions obtained for $\icw=2$, $\icw=8$ and $\icw=12$ were
  indistinguishable from those shown in~\fig{fh2b-prof}.
These regimes are similar to those described
in~\cite{Tsyganov-etal-2003}.

The behaviour for $\fna=0.07$ is different: the shape of
the pulse changes as it propagates. These are ``envelope
quasisolitons'', similar to those previously reported in
in~\cite{Biktashev-Tsyganov-2011,Tsyganov-Biktashev-2014}, and can be
described as modulated high-frequency waves with the envelope in the
form of a solitary wave, where the speed of the high-frequency waves
(the ``phase velocity'') is different from the speed of the envelope
(the ``group velocity''), thus the change in shape. Note that this
behaviour is similar to solitons in the nonlinear
Schr\"odinger equation (NLS)~\cite{NLS},
both in terms of the varying shape and the
reflection from boundary, with the difference that here the system is
dissipative so here again
there are unique and stable amplitude and envelope
shape, with corresponding group and phase velocities, which are
determined by the parameters of the system but do not depend on
initial conditions, as long as a propagating wave is initiated.

Since the speed and shape of quasi-solitons are
  typically fixed, the reflections and collisions, when they happen,
  are always quasi-elastic, if one is to use the terminology from the
  conservative wave theory. That is, the properties of the solitons
  fully recover after collisions, even though, as can be seen
  in~\fig{fhdp}, it make take some time. On the other hand, there are
  examples of quasi-solitons models in which the unique shape of
  quasi-solitons takes a very long time to establish, so at short time
  intervals, one may consider a one-parametric family of quasi-soliton
  solutions: in~\cite{Tsyganov-Biktashev-2004}, these are solutions
  differing in their ``thickness'', i.e. distance between the front
  and the back. The rules of collision there are more complicated,
  e.g. there could be ``completely inelastic'' collisions, where of the two
  colliding waves one survives and the other annihilates.

Finally, we emphasize that the reflection from
  boundaries shown in~\fig{fhdp} is for homogeneous Neumann boundary
  condition. Replacing those with e.g. homogeneous Dirichlet boundary
  conditions makes reflection much less likely. In particular, it is
  not observed in any of the six cases shown in \figs{fh2b-prof} and
  \figref{fhdp}, although can be observed for other parameter values,
  for instance for $\fna=0.03$, and even then the reflected wave takes
  much longer to fully recover. In this aspect, the quasi-solitons are
  also different from the true solitons, e.g. in the NLS. A formal way
  to understand this difference is to observe that NLS is symmetric
  with respect to inversion of the sign of its complex field. Hence
  one can arrange two identical but counter-propagating solitons on an
  infinite line so that their nonlinear superposition at a certain
  point remains exactly zero at all times. Then replacing the problem
  with the one at half-line and zero boundary condition at that point
  will yield a solution in the form of a soliton reflecting from the
  boundary with a $180^\deg$ change of phase. This construction does
  not work for the FitzHugh-Nagumo kinetics~\eq{FHN} which is not
  invariant with respect to the inversion $(\u,\v)\to(-\u,-\v)$,
  unless $\fna=-1$, but in the latter case the system is no longer
  excitable. A more intuitive way to explain this is: for the pulse
  way to propagate, the $\u$ field must exceed the threshold $\fna$,
  and the boundary condition $\u=0$ makes it much more difficult for
  the reflected wave to satisfy this.

\section{Discussion}
\seclabel{Discussion}
It has been traditionally believed that a definitive property of waves in excitable media is that they annihilate when collided. Although soliton-like interaction was observed in some reaction-diffusion systems with excitable kinetics, both in numerical simulations~\cite{%
  Petrov-etal-1994,%
  Aslanidi-Mornev-1997%
} and in experiments~\cite{%
  vonOertzen-etal-1998,%
  Tsyganov-etal-1993%
}, solitons are mostly studied in fully integrable systems (KdV, sin-Gordon, nonlinear Schr\"odinger).  Perhaps the main reason of this view on excitable media was that soliton-like interactions were always limited to narrow parameter ranges close to the boundaries between excitable and oscillatory (limit cycle) regimes of the reaction kinetics.  
A crucial role in the change of the attitude to excitable media as a
source of solitons was played by experimental and theoretical works by
Vanag and Epstein~\cite{%
  Vanag-Epstein-2002,%
  Vanag-Epstein-2004,%
  Vanag-Epstein-2009,%
  Vanag-book%
}. They have demonstrated reaction-diffusion systems with soliton-like
interaction of waves, and also spontaneous formation of wave
packets. At the same time, we have shown that in excitable systems
with cross-diffusion, the soliton-like behaviour of waves can be quite
typical, including solutions similar to group (envelope)
solitons~\cite{%
  Biktashev-Tsyganov-2011%
}. These works resonate with Vanag and Epstein's reports of
cross-diffusion in the chemical system BZ-AOT~\cite{%
  Vanag-Epstein-2009%
}. 
In the present work, we have demonstrated that quasi-solitons,
including group (envelope) quasi-solitons, can observed in
reaction-diffusion systems with self-diffusion only. 
This has been found with the help of 
two-component systems with effective cross-diffusion, which are
obtained by semi-rigorous adiabatric reduction of a multicomonent
reaction-diffusion system with self-diffusion only. 
Adiabatic elimination of fast fields, which gives rise to
  nontrivial dissipative terms, that, for physical reasons, cannot
  exist in the straightforward form, can be observed in various physical
  settings. For instance, in nonlinear optics, adiabatic elimination of
  the acoustic field gives rise to an extra term in the NLS equation
  for the optical field, that is similar to stimulated Raman
  scattering which cannot appear in that equation directly, thus
  dubbed ``pseudo-stimulated-Raman-scattering''~\cite{%
    Gromov-Malomed-2013-JPP,%
    Gromov-Malomed-2014-OC,%
    Gromov-Malomed-2015-PRE%
    }. By the analogy with that result, the effective cross-diffusion
    we described here could be called ``pseudo-cross-diffusion''.
    For the purposes of the present communication, an important
feature of the effective cross-diffusion is that the resulting
diffusivity matrix is \emph{not} constraint by the thermodynamic restrictions
of symmetry and positive definiteness. We believe that application of
such reduction, accounting for the emergence of effective
cross-diffusion, may lead to finding new interesting regimes in
systems that have traditionally been studied without cross-diffusion,
e.g. Brusselator~\cite{%
  Tyson-Light-1973%
} and Oregonator~\cite{%
  Field-Noyes-1974,%
  Field-Burger-1985%
}, which remains an interesting direction for further study.

\subsection*{Acknowledgements}
%
VNB is supported in part by EPSRC grant EP/N014391/1 (UK). 

\subsection*{Authors contribution}
Authors made contributions of equal value to this work. VNB mostly
contributed to the analytical part and MAT mostly contributed to the
numerical part. 

\subsection*{Competing financial interests}
Authors do not have any competing financial interests. 

\subsection*{Correspondence}
Please address all correspondence to: V.N.~Biktashev, College of
Engineering, Mathematics and Physical Sciences, University of Exeter,
Exeter EX4 4QF, UK, email~\url{v.n.biktashev@exeter.ac.uk}.

\bibliographystyle{unsrt}

\end{document}